\documentclass[reprint,aps,prl,twocolumn,superscriptaddress,
longbibliography,floatfix,secnumarabic,footinbib]{revtex4-1}%
\usepackage{graphicx}
\usepackage{amsfonts}
\usepackage{amsmath}
\usepackage{ltxtable}
\usepackage{CJK}
\usepackage{mdwmath}
\usepackage{color}
\usepackage{appendix}
\usepackage{mathtools}
\usepackage{amssymb}
\usepackage{bm}
\usepackage{bbold}
\usepackage{booktabs}
\usepackage{threeparttable}
\usepackage[normalem]{ulem}
\usepackage{makecell}
\usepackage[colorlinks,allcolors=blue]{hyperref}

\newcommand{\mI}{{\mathcal I}}
\newcommand{\mB}{{\mathcal B}}

\newcommand{\mS}{{\mathcal S}}

\begin{document}

\title{Bell Inequalities Induced by Pseudo Pauli Operators on Single Logical Qubits}

\author{Weidong Tang}
\email{wdtang@snnu.edu.cn}
\affiliation{School of Mathematics and Statistics, Shaanxi Normal University, Xi'an 710119, China}

\begin{abstract}

In most Bell tests, the measurement settings are specially chosen so that the maximal quantum violations of the  Bell inequalities
can be detected, or at least, the violations are strong enough to be observed.
Such choices can usually associate the corresponding Bell operators
to a kind of effective observables, called pseudo Pauli operators, providing us a more intuitive understanding of Bell nonlocality in some sense.
Based on that,  a more general quantum-to-classical approach for the constructions of  Bell inequalities is developed. Using
this approach, one can not only derive several kinds of well-known Bell inequalities, but also explore many new ones.
Besides, we show that some quadratic Bell inequalities can be induced from the uncertainty relations of pseudo Pauli operators as well,
which may shed new light on the study of uncertainty relations of some nonlocal observables.

\end{abstract}
\maketitle

Some entangled states shared by distant observers may exhibit a counterintuitive feature called the Bell nonlocality, i.e.,
correlations produced by measurements on space-like separated subsystems cannot be simulated by any local hidden variable (LHV) model.
This non-classical feature can usually be shown by the violation of a Bell inequality. So far, various Bell inequalities have been proposed,
including the original version\cite{Bell1964},
the Clauser-Horne-Shimony-Holt (CHSH) inequality\cite{CHSH1969}, the Mermin inequality\cite{Mermin1990},  the Collins-Gisin-Linden-Massar-Popescu inequality\cite{CGLMP2002} and etc. These Bell inequalities are widely used in many quantum information tasks, such as quantum games\cite{Cleve2004,Linden2007,Pappa2015}, device-independent quantum key distribution\cite{Acin2007,Franz2011,Masanes2011}, random number generation\cite{Pironio2010,Acin2012}, and self-testing\cite{McKague2012,Yang2013,Supic2016,Baccari2020}.

From an overall perspective, there are mainly two ways to construct Bell inequalities. The first one is known as
the classical-to-quantum approach (or the local polytope approach\cite{Froissart1981,Garg-Mermin1984,Peres1999,Brunner2014}), which is also regarded as the standard construction of Bell inequalities. It exploits a mathematical tool called the (local) polytope, which is the convex hull of a finite number of vertices, to represent  the set of correlations admitting a local hidden variable model,
where the vertices correspond to local deterministic assignments (using 1 and 0 to describe the corresponding deterministic behaviors)\cite{Brunner2014}, and the facets can define a finite set of linear inequalities which are called facet (or tight) Bell inequalities. A facet inequality is nontrivial if it can be violated on
some quantum systems. Finding such facet inequalities is
the central task in this approach.
The constructions of the Collins-Gisin-Linden-Massar-Popescu inequality\cite{CGLMP2002}, the Froissard inequality\cite{Froissart1981},
and the Collins-Gisin inequality\cite{Collins2004} are all of this type.
The main shortcoming of this approach is the lack of efficiency, especially in the scenarios
involving a large number of parties (or measurements per party, or outcomes for each measurement).
By contrast, the second way,  also known as the quantum-to-classical approach\cite{Lim2010,Salavrakos2017,Baccari2020}, seems to be more practical in constructing scalable Bell inequalities. The key of this approach relies on how to properly exploit quantum properties of the states with special symmetries. Therefore, the involved states are usually chosen from the stabilizer states, and
their stabilizers would be invoked directly\cite{Cabello2008-1,Tang2013,Tang2017-1} or indirectly\cite{Lim2010,Baccari2020} in the constructions of Bell inequalities.

Although the violation of the Bell inequality can be used for revealing some quantum features such as the nonlocality and entanglement,
the inequality itself looks more like a kind of mathematical tool since it is essentially a constraint of classical correlations.
The explorations on more explicit physical significance for the Bell inequality are still very rare so far. In view of this, one motivation of this work is to
give a new physical explanation for some (e.g. the maximal) quantum violations of certain Bell inequalities. To realize that, a natural thought is to look for
some special connections between the Bell operators and certain physical quantities.
Once such connections are found, they might shed new light on the constructions of Bell inequalities. Then,
to explore a new approach for the constructions of Bell inequalities is another  motivation of this work.

To start, let us consider the two-qubit scenario since it is
a stepping stone to many multi-qubit generalizations, and in which the typical Bell inequality is the CHSH inequality. Below we will show how to associate the corresponding Bell operator to a physical quantity, and conversely, how to derive the CHSH inequality from such a physical quantity.

Denote by $X,Y,Z$ the Pauli matrices $\sigma_x,\sigma_y,\sigma_z$, and let $|0\rangle,|1\rangle$ be the eigenstates of $Z$ with eigenvalues $+1,-1$ respectively.
Consider a two-dimensional Hilbert space spanned by two Bell states
 \begin{align}\label{logical-qubit-basis}
    |\tilde{0}\rangle=\frac{1}{\sqrt{2}}(|00\rangle+|11\rangle),~
    |\tilde{1}\rangle=\frac{1}{\sqrt{2}}(|01\rangle-|10\rangle).
\end{align}
Recall that a logical qubit can be encoded into
several physical qubits. In view of this,  the state
\begin{align}\label{logical-qubit-def}
 |\tilde{\psi}\rangle=\alpha|\tilde{0}\rangle+&\beta|\tilde{1}\rangle~(|\alpha|^2+|\beta|^2=1),
\end{align}
can be regarded as a special logical qubit, and $\{|\tilde{0}\rangle,|\tilde{1}\rangle\}$ is also called a logical qubit basis.

Besides, one can define three {\it pseudo Pauli operators} on such a logical qubit as follows:
\begin{align}\label{logical-Pauli-XYZ}
      \tilde{Z}=&|\tilde{0}\rangle\langle\tilde{0}|-|\tilde{1}\rangle\langle\tilde{1}|=\frac{1}{2}(Z_1Z_2+X_1X_2),\cr
      \tilde{X}=&|\tilde{0}\rangle\langle\tilde{1}|+|\tilde{1}\rangle\langle\tilde{0}|=\frac{1}{2}(Z_1X_2-X_1Z_2),\cr
      \tilde{Y}=&i(|\tilde{1}\rangle\langle\tilde{0}|-|\tilde{0}\rangle\langle\tilde{1}|)=\frac{1}{2}(Y_2-Y_1).
\end{align}
Clearly, their eigenvalues are all $\pm1$, the same as those of common Pauli operators. Apart from this definition,  Appendix A also shows us several examples based on other logical qubit bases.

On the other hand, it is known that the CHSH inequality can be written as
\begin{align}\label{CHSH-ori}
 |\langle\mB_{\text{CHSH}}\rangle_c|=|\langle A_1B_2+A_1B_2^{\prime}+A_1^{\prime}B_2-A_1^{\prime}B_2^{\prime}\rangle_c|\leq2,
\end{align}
where $\langle~\rangle_c$ denotes the classical expectation (by the LHV model), and
 \begin{align}\label{Bell-op-ori}
   \mB_{\text{CHSH}} \equiv  A_1B_2+A_1B_2^{\prime}+A_1^{\prime}B_2-A_1^{\prime}B_2^{\prime}.
\end{align}
is the corresponding Bell operator. Here $A_1,A_1^{\prime},B_2,B_2^{\prime}$ are dichotomic observables measured by two distant observers.
As is known, its quantum expectation satisfies $\langle\mB_{\text{CHSH}}\rangle\leq2\sqrt{2}$, i.e.,  the
maximal quantum violation of the CHSH inequality is $2\sqrt{2}$. This bound can only be attained by
the maximally entangled states. For example, a common choice is the Bell state $|\tilde{0}\rangle=(|00\rangle+|11\rangle)/\sqrt{2}$.
Accordingly, the observables to be measured are  $A_1=X_1,A_1^{\prime}=Z_1,B_2=(X_2+Z_2)/\sqrt{2}$ and $B_2^{\prime}=(X_2-Z_2)/\sqrt{2}$, i.e.,  the Bell operator to be tested in the real experiment (will referred to as experimental Bell operator hereinafter) is
\begin{align}\label{Bell-op-real-exp}
   \mB_{\text{CHSH}}^{\text{e}}=&X_1\frac{X_2+Z_2}{\sqrt{2}}+ X_1\frac{X_2-Z_2}{\sqrt{2}}\cr
   &+ Z_1 \frac{X_2+Z_2}{\sqrt{2}}- Z_1 \frac{X_2-Z_2}{\sqrt{2}}.
\end{align}
Note that the observables to be measured on each observer's side are a pair of complementary  observables (e.g. $X$ and $Z$), which is the price of testing the maximal violation of the CHSH inequality.

Rewrite $\mB_{\text{CHSH}}^{\text{e}}$ in terms of a combination of two stabilizers of $(|00\rangle+|11\rangle)/\sqrt{2}$, i.e.,
\begin{align}\label{CHSH-op-stabilizer-rep}
    \mB_{\text{CHSH}}^{\text{e}}=\sqrt{2}(X_1X_2+Z_1Z_2)\equiv\mB_{\text{CHSH}}^{\text{e}}(S),
\end{align}
where  $X_1X_2$ and $Z_1Z_2$ are the stabilizers. Besides, according to Eq.(\ref{logical-Pauli-XYZ}) and Eq.(\ref{CHSH-op-stabilizer-rep}), $\mB_{\text{CHSH}}^{\text{e}}$ can be further represented as
\begin{align}\label{Bell-op-simplest}
    \mB_{\text{CHSH}}^{\text{e}}=2\sqrt{2}\tilde{Z}\equiv\mB_{\text{CHSH}}^{\text{e}}(L),
\end{align}
which corresponds to an explicit physical quantity on a single logical qubit!  This quantity
can be considered as the $z$-component of an pseudo spin (up to a constant).
In fact, choosing other maximally entangled states and proper measurement settings will induce similar correspondences as well.

On the other hand,  reversing the above discussion, i.e., $\mB_{\text{CHSH}}^{\text{e}}(L)\rightarrow\mB_{\text{CHSH}}^{\text{e}}(S)\rightarrow\mB_{\text{CHSH}}^{\text{e}}
\rightarrow\mB_{\text{CHSH}}$, one can construct a CHSH inequality.
Note that the expression of $\mB_{\text{CHSH}}^{\text{e}}(L)$ (or $\mB_{\text{CHSH}}^{\text{e}}(S)$) indicates that
the quantum upper bound of $\mB$ is at least $2\sqrt{2}$. The bound derived directly from $\mB_{\text{CHSH}}^{\text{e}}(L)$  is a rough estimation for
the maximal quantum violation (will be referred to as rough quantum upper bound below) of the final Bell inequality.
To show that $2\sqrt{2}$ is also the exact quantum upper bound,
one can use a sum of square method\cite{Supic2016,Salavrakos2017,Baccari2020}, see Appendix B. By contrast, the classical upper bound can be derived by invoking the (classical) inequality $|x+y+z-xyz|\leq2~(-1\leq x,y,z\leq1)$. Since $2<2\sqrt{2}$, such a construction is successful.

Inspired by that, we can propose a systematic (quantum-to-classical) approach to construct Bell inequalities. It can be simply described by the sequence
\begin{align}\label{Bell-sequence}
    \mB^{\text{e}}(L)\rightarrow\mB^{\text{e}}(S)\rightarrow\mB^{\text{e}}\rightarrow\mB,
\end{align}
and an  additional comparison for the classical upper bound and the quantum one of $\mB$. As mentioned above, a rough quantum upper bound can be given directly by
$\mB^{\text{e}}(L)$. In the demonstration of Bell nonlocality,
if the gap between this bound and the classical one is large enough, one do not really have to calculate the exact quantum upper bound (except for some specific tasks).
Fortunately, for many famous Bell inequalities, their exact quantum upper bounds can be calculated by a sum of square method, see Appendix B.
If the classical upper  bound is less than the (rough or exact) quantum one, a desired Bell inequality is constructed; otherwise, choose another $\mB^{\text{e}}(L)$ and repeat the above process. Note that from the matrix perspective,
$\mB^{\text{e}}(L)=\mB^{\text{e}}(S)=\mB^{\text{e}}$, but their physical significance may be different.

We prefer to choose two fully entangled stabilizer states (such as Bell states and some graph states) as logical qubit bases in constructing Bell inequalities,
since one can easily detect quantum violations for many celebrated Bell inequalities
in such states, and besides,  the preparations of stabilizer states
are more attainable than non-stabilizer ones  by current experiments.

Usually we can choose $\mB^{\text{e}}(L)=\beta_q\vec{k}\cdot \vec{\tilde{\sigma}}$, where $\beta_q$ is a constant (rough quantum upper bound),
$\vec{k}$ is a unit vector and $\vec{\tilde{\sigma}}\equiv(\tilde{X},\tilde{Y},\tilde{Z})$. Besides, the expressions of
$\tilde{X},\tilde{Y},\tilde{Z}$ invoked in $\mB^{\text{e}}(S)$ rely on the choice of the logical qubit basis,
and each of them contains $2^{n-1}$ terms (since the bases are fully entangled stabilizer states, see Appendix B),
 where $n$ is the number of physical qubits in each basis.
Note that if $\mB^{\text{e}}(L)$ is properly chosen, $\mB^{\text{e}}$ and $\mB^{\text{e}}(S)$ might take the same form.
Apart from that, another possible form of
$\mB^{\text{e}}$ could be given by decomposing each term of $\mB^{\text{e}}(S)$ into a pair of complementary  observables
(e.g  from Eq.(\ref{CHSH-op-stabilizer-rep}) to Eq.(\ref{Bell-op-real-exp})). Since the algebraic structures of $\mB^{\text{e}}$
and  $\mB$ are the same, one can easily get the latter by applying a suitable replacement of operators.

\begin{table*}\caption{New constructions of three typical Bell inequalities. } \label{TB1}
    \centering
    \begin{tabular}{c|ccc}
    \hline
    \hline
        & CHSH  & Mermin & Svetlichny   \\
  \hline
     \parbox[c]{2.2cm}{$|\tilde{0}\rangle$ }  & $\frac{1}{\sqrt{2}}(|00\rangle+|11\rangle)$ &  $\frac{1}{2}[|0\rangle(|00\rangle-|11\rangle)-|1\rangle(|01\rangle+|10\rangle)]$ &  $\frac{1}{2}[|0\rangle(|00\rangle-|11\rangle)-|1\rangle(|01\rangle+|10\rangle)]$  \\
     \parbox[c]{2.2cm}{$|\tilde{1}\rangle$ }  & $\frac{1}{\sqrt{2}}(|01\rangle-|10\rangle)$ &
     $\frac{1}{2}[|0\rangle(|01\rangle+|10\rangle)+|1\rangle(|00\rangle-|11\rangle)]$ & $\frac{1}{2}[|0\rangle(|01\rangle+|10\rangle)+|1\rangle(|00\rangle-|11\rangle)]$ \\
     \parbox[c]{2.2cm}{$\tilde{X}$ }  & $\frac{1}{2}(Z_1X_2-X_1Z_2)$ & \thead{$\frac{1}{4}(X_1Z_2Z_3+Z_1X_2Z_3$ \\ $+Z_1Z_2X_3-X_1X_2X_3)$} & \thead{$\frac{1}{4}(X_1Z_2Z_3+Z_1X_2Z_3$ \\ $+Z_1Z_2X_3-X_1X_2X_3)$} \\
     \parbox[c]{2.2cm}{$\tilde{Z}$ }  & $\frac{1}{2}(X_1X_2+Z_1Z_2)$ & \thead{$\frac{1}{4}(Z_1Z_2Z_3-Z_1X_2X_3$ \\ $-X_1Z_2X_3-X_1X_2Z_3)$} & \thead{$\frac{1}{4}(Z_1Z_2Z_3-Z_1X_2X_3$ \\ $-X_1Z_2X_3-X_1X_2Z_3)$} \\
      \parbox[c]{2.2cm}{ $\mB^{e}(L)$ }  &  $2\sqrt{2}\tilde{Z}$ or $2\sqrt{2}\frac{\tilde{Z}+\tilde{X}}{\sqrt{2}}$ & $4\tilde{Z}$ & $4\sqrt{2}\tilde{Z}$ or $4\sqrt{2}\frac{\tilde{X}-\tilde{Z}}{\sqrt{2}}$ \\
      \parbox[c]{2.2cm}{ $\mB$} & \thead{$A_1 B_2+A_1B_2^{\prime}$\\ $+A_1^{\prime}B_2-A_1^{\prime}B_2^{\prime}$}         & \thead{$A_1A_2A_3-A_1B_2B_3$\\ $-B_1A_2B_3-B_1B_2A_3$}  & \thead{$A_1B_2B_3+B_1A_2B_3+B_1B_2A_3-A_1A_2A_3$\\ $+B_1A_2A_3+A_1B_2A_3+A_1A_2B_3-B_1B_2B_3$} \\
         \parbox[c]{2.2cm}{ Bell inequality }  & $|\langle\mB\rangle_c|\leq2$ & $|\langle\mB\rangle_c|\leq2$ & $|\langle\mB\rangle_c|\leq4$\\
         \parbox[c]{2.2cm}{ $\langle\mB\rangle_{\max}$ }  & $2\sqrt{2}$ & $4$ & $4\sqrt{2}$\\
     \hline
     \hline
    \end{tabular}
\end{table*}

{\it Example 1.} --- The CHSH inequality, the three-qubit Mermin inequality and the three-qubit Svetlichny inequality. According to Eq.(\ref{Bell-sequence}), to construct any of them,
we need to choose a suitable logical qubit basis and a proper $\mB^{\text{e}}(L)$, which are listed in Table \ref{TB1}, also see Appendix C for more detailed discussions.
As mentioned above, their maximal quantum violations $\langle\mB\rangle_{\max}$  can be derived by the sum of square
method.
\hfill $\blacksquare$

{\it Example 2.} --- Graphical Bell inequalities induced from the  quantum error-correcting code $[[5,1,3]]$\cite{Bennett1996}.
Let $|\tilde{0}\rangle=|L_{5}\rangle$ and $|\tilde{1}\rangle=Z_0Z_1Z_2Z_3Z_4|L_{5}\rangle$ be the bases,
which are also the bases for the coding space, where $|L_{5}\rangle$ is
a $5$-vertex loop graph state, which is stabilized by $g_{i\oplus1}=Z_{i}X_{i\oplus1}Z_{i\oplus2}~(i=0,1,\cdots,4)$,
and $\oplus$ stands for addition modulo 5.
Note that $|L_{5}\rangle\langle L_{5}|=\prod_{i=0}^{4}\frac{I+g_i}{2}$. Then the pseudo Pauli operators can be written as
\begin{align}\label{L5-Z-X-Y}
    \tilde{Z}
    =&\frac{1}{16}[\sum_{i=0}^4(Z_{i}X_{i\oplus1}Z_{i\oplus2}-Z_iY_{i\oplus1}X_{i\oplus2}Y_{i\oplus3}Z_{i\oplus4}\cr
    &+X_iY_{i\oplus2}Y_{i\oplus3})-X_0X_1X_2X_3X_4],\cr
    \tilde{X}
    =&\frac{1}{16}[\sum_{i=0}^4(-Y_iZ_{i\oplus1}Y_{i\oplus2}+Y_iX_{i\oplus1}Z_{i\oplus2}X_{i\oplus3}Y_{i\oplus4}\cr
    &-Z_iX_{i\oplus2}X_{i\oplus3})+Z_0Z_1Z_2Z_3Z_4],\cr
     \tilde{Y}
    =&\frac{1}{16}[\sum_{i=0}^4(-X_iY_{i\oplus1}X_{i\oplus2}+X_iZ_{i\oplus1}Y_{i\oplus2}Z_{i\oplus3}X_{i\oplus4}\cr
    &-Y_iZ_{i\oplus2}Z_{i\oplus3})+Y_0Y_1Y_2Y_3Y_4].\cr
\end{align}
Detailed calculations are shown in Appendix D. There are three typical choices for Bell operators,
\begin{align}\label{graphical-Bell-op-simp-L}
\mB^e_{1}(L)=16\tilde{Z};&~~~
\mB^e_{2}(L)=16\sqrt{2}\frac{\tilde{Z}+\tilde{X}}{\sqrt{2}};\cr
\mB^e_{3}(L)=&16\sqrt{3}\frac{\tilde{Z}+\tilde{X}+\tilde{Y}}{\sqrt{3}}.
\end{align}
They can induce three Bell inequalities: the Mermin-like inequality, the Svetlichny-like inequality, and the Hyper-Svetlichny-like inequality. Numerical calculations show that their classical bounds satisfy
\begin{align}\label{graphical-Bell-ineq}
   |\langle\mB_1\rangle_c|\leq8;~~|\langle\mB_2\rangle_c|\leq16;~~|\langle\mB_3\rangle_c|\leq24.
\end{align}
Here $\mB_{1,2,3}$ are final Bell operators. One can get them by substituting the expressions of pseudo Pauli operators into Eq.(\ref{graphical-Bell-op-simp-L}) and replacing $Z_i,X_i,Y_i$ with $A_i,B_i,C_i$ respectively. For example, $\mB_1=\sum_{i=0}^4(A_{i}B_{i\oplus1}A_{i\oplus2}-A_iC_{i\oplus1}B_{i\oplus2}C_{i\oplus3}A_{i\oplus4} +B_iC_{i\oplus2}C_{i\oplus3})-B_0B_1B_2B_3B_4$, where $A_i,B_i,C_i\in[-1,1]$.

By contrast, the quantum upper bound for the Mermin-like inequality can be easily derived. But for the other two, we
we only give their rough quantum upper bounds by exploiting Eq.(\ref{graphical-Bell-op-simp-L}), namely,
\begin{align}\label{graphical-Bell-ineq}
   \langle\mB_1\rangle_{\max}=16;~~\langle\mB_2\rangle_{\max}\geq16\sqrt{2};~~\langle\mB_3\rangle_{\max}\geq16\sqrt{3}.
\end{align}
Clearly, the gap between any above (exact or rough) quantum upper bound and the corresponding classical one
is large enough for the detection of Bell nonlocality.  This example can be generalized to the $(2k+1)$-qubit scenario ($k\geq2$) as well.
\hfill $\blacksquare$

{\it Example 3.} --- The chained (or multi-setting) Bell inequality\cite{Braunstein1990}.
We still adopt the definition of $|\tilde{0}\rangle$, $|\tilde{1}\rangle$, $\tilde{Z}$ and $\tilde{X}$ by Eq.(\ref{logical-qubit-basis}) and Eq.(\ref{logical-Pauli-XYZ}). Denote $Z_i(\theta):=Z_i\cos\theta+X_i\sin\theta$, where $i\in\{1,2\}$ and $\theta\in[0,\pi)$.
One can  rewrite $\tilde{Z}$ in Eq.(\ref{logical-Pauli-XYZ}) into another form:
\begin{align}\label{Z-t-another-rep}
\tilde{Z}=\frac{1}{n}\sum_{k=1}^{n}Z_1(\frac{k-1}{n}\pi)Z_2(\frac{k-1}{n}\pi);~n\geq2.
\end{align}
Besides, $\tilde{X}=\tilde{Z}\cdot (iY_2)=\frac{1}{n}\sum_{k=1}^{n}Z_1(\frac{k-1}{n}\pi)Z_2(\frac{n+2k-2}{2n}\pi)$. We can choose
\begin{align}\label{Chained-Bell-op-simp-L}
\mB^e_C(L)=2n\cos\frac{\pi}{2n}\tilde{Z}.
\end{align}
Then $\mB^e_C(S)=2\cos\frac{\pi}{2n}\sum_{k=1}^{n}Z_1(\frac{k-1}{n}\pi)Z_2(\frac{k-1}{n}\pi)$ and
$\mB^e_C=\sum_{k=1}^{n}Z_1(\frac{k-1}{n}\pi)Z_2(\frac{2k-1}{2n}\pi)+\sum_{k=1}^{n-1}Z_1(\frac{k}{n}\pi)Z_2(\frac{2k-1}{2n}\pi)-Z_1(0)\otimes Z_2(\frac{2n-1}{2n}\pi)$. Here we have exploited such a relation:  for each $2\leq k\leq n$, $Z_i(\frac{k-1}{n}\pi)=[Z_i(\frac{2k-3}{2n}\pi)+Z_i(\frac{2k-1}{2n}\pi)]/\cos\frac{\pi}{2n}$, and for $k=1$,$Z_i(0)=[Z_i(\frac{1}{2n}\pi)-Z_i(\frac{2n-1}{2n}\pi)]/\cos\frac{\pi}{2n}$.

Replacing $Z_1(\frac{k-1}{n}\pi)$ and $Z_2(\frac{2k-1}{2n}\pi)$ with $A_1^k$ and $B_2^k$ respectively, one can get the final Bell operator
\begin{align}\label{Chained-Bell-op}
\mB_C=\sum_{k=1}^{n}A_k\otimes B_k+\sum_{k=1}^{n-1}A_{k+1}\otimes B_k-A_1\otimes B_n.
\end{align}
Notice that $|c_1+c_2+\cdots+c_{2n-1}-\prod_{i=1}^{2n-1}c_i|\leq2n-2$ ($\forall i\in\{1,2,\cdots,2n-1\}$, $c_i\in[-1,1]$). Therefore,
\begin{align}\label{Chained-Bell-eq}
|\langle\mB_C\rangle_c|\leq2n-2.
\end{align}
According to Eq.(\ref{Chained-Bell-op-simp-L}), its quantum upper bound is at least $2n\cos\frac{\pi}{2n}$, which is also the exact bound.
One can prove that by using the sum of square method, see Ref.\cite{Supic2016}. since $2n\cos\frac{\pi}{2n}>2n[1-\pi^2/(8n^2)]>2n-2$,
Eq.(\ref{Chained-Bell-eq}) is a desired Bell inequality.
\hfill $\blacksquare$

To construct a multi-partite Bell inequality, apart from a direct choice of stabilizer states as logical qubit bases, another way
is to use a special recursive method, which is also a generalization of Mermin-Klyshko polynomial technique\cite{Mermin1990,Klyshko1993,Belinski1993}. To show that, first we introduce a linear expansion operation $\ast$:
\begin{align*}
 (a)~ &\ast(\alpha|0\rangle+\beta|1\rangle)=\alpha \ast(|0\rangle)+\beta \ast(|1\rangle)=\alpha|\tilde{0}\rangle+\beta|\tilde{1}\rangle;\cr
 (b)~ &\ast(u X+v Y +w Z)=u \ast(X)+v \ast(Y) +w \ast(Z)\cr
      &=u \tilde{X}+v \tilde{Y} +w \tilde{Z}.
\end{align*}
where $u,v,w$ are real numbers. Besides, $|0\rangle$ ($|\tilde{0}\rangle$) and $|1\rangle$ ($|\tilde{1}\rangle$) are physical (logical) qubits, while
$X,Y,Z$ ($\tilde{X},\tilde{Y},\tilde{Z}$) are normal (pseudo) Pauli operators.
Denote by  $\ast_k$ the expansion operation applied to the state or the Pauli operators on the $k$-th physical qubit.

For simplicity, here we only discuss a special iteration.  Let the initial basis and  pseudo Pauli operators be $|\tilde{0}\rangle^{(1)}=|0\rangle$, $|\tilde{1}\rangle^{(1)}=|1\rangle$, $\tilde{X}^{(1)}=X$ and $\tilde{Z}^{(1)}=Z$. The iteration can be described as follows: $|\tilde{0}\rangle^{(n)}=\ast_{n-1}(|\tilde{0}\rangle^{(n-1)})$,
$|\tilde{1}\rangle^{(n)}=\ast_{n-1}(|\tilde{1}\rangle^{(n-1)})$, $\tilde{X}^{(n)}=\ast_{n-1}(\tilde{X}^{(n-1)})$ and $\tilde{Z}^{(n)}=\ast_{n-1}(\tilde{Z}^{(n-1)})$, i.e., $\ast$ always acts on the last qubit in each round.
Besides, one can also prove that
\begin{align}\label{general-logical-X-Z-def}
\tilde{Z}^{(n)}=&(|\tilde{0}\rangle\langle\tilde{0}|)^{(n)}-(|\tilde{1}\rangle\langle\tilde{1}|)^{(n)},\cr \tilde{X}^{(n)}=&(|\tilde{0}\rangle\langle\tilde{1}|)^{(n)}+(|\tilde{1}\rangle\langle\tilde{0}|)^{(n)}.
\end{align}
Then choosing a suitable $\mB^{\text{e}}_n(L)$ (e.g. $\mB^{\text{e}}_n(L)=2^{n-1}\tilde{Z}^{(n)}$), and following the steps given by Eq.(\ref{Bell-sequence}), one can derive the desired Bell inequality.

For the two-qubit scenario, invoking a special definition of $|\tilde{0}\rangle$ and $|\tilde{1}\rangle$ from Eq.(\ref{logical-qubit-basis}), one have $|\tilde{0}\rangle^{(2)}=\ast_{1}|0\rangle=|\tilde{0}\rangle=(|00\rangle+|11\rangle)/\sqrt{2}$,
$|\tilde{1}\rangle^{(2)}=\ast_{1}|1\rangle=|\tilde{1}\rangle=(|01\rangle-|10\rangle)/\sqrt{2}$,
$\tilde{X}^{(2)}=(Z_1Z_2+X_1X_2)/2$ and $\tilde{Z}^{(2)}=(Z_1X_2-X_1Z_2)/2$.
 Choosing $\mB^{\text{e}}_2(L)=2\sqrt{2}\tilde{Z}^{(2)}$, one can derive the CHSH Bell inequality, see Table \ref{TB1}.

For the three-qubit scenario,  by applying $\ast_2$ to $|\tilde{0}\rangle^{(2)}$, $|\tilde{1}\rangle^{(2)}$, $\tilde{Z}^{(2)}$ and $\tilde{X}^{(2)}$, one can obtain
\begin{align}\label{star2-operation}
&|\tilde{0}\rangle^{(3)}=\frac{1}{2}(|000\rangle+|011\rangle+|101\rangle-|110\rangle),\cr
&|\tilde{1}\rangle^{(3)}=\frac{1}{2}(|001\rangle-|010\rangle-|100\rangle-|111\rangle);\cr
&\tilde{Z}^{(3)}=\frac{1}{4}(Z_1Z_2Z_3+Z_1X_2X_3+X_1Z_2X_3-X_1X_2Z_3),\cr
&\tilde{X}^{(3)}=\frac{1}{4}(Z_1Z_2X_3-Z_1X_2Z_3-X_1Z_2Z_3-X_1X_2X_3).\cr
\end{align}
Likewise, choosing $\mB^{\text{e}}_3(L)=4\tilde{Z}^{(3)}$ will give rise to $\mB^{\text{e}}_3=Z_1Z_2Z_3+Z_1X_2X_3+X_1Z_2X_3-X_1X_2Z_3$. Accordingly, the Bell inequality is $|\langle\mB_3\rangle_c|\leq2$, where $\mB_3=A_1A_2A_3+A_1B_2B_3+B_1A_2B_3-B_1B_2A_3$.  It is a three-qubit Mermin inequality (up to a local unitary transformation).
Table \ref{TB1} provides us a similar construction based on another choice of $|\tilde{0}\rangle$ and $|\tilde{1}\rangle$. Clearly, this technique applies to the construction of the Svetlichny inequality as well, also see Table \ref{TB1}.

Repeating the above recursive operations can give the $n$-qubit Mermin-type and Svetlichny-type inequalities, where $n\geq3$, see Appendix E.
Similar iterations can apply to the constructions of multi-partite chained-like Bell inequalities, various hybrid Bell inequalities(e.g. chained and Mermin-like  Bell inequalities) and etc.

Finally, we would like to discuss the connections between the pseudo Pauli operators and some quadratic Bell inequalities.
Quadratic Bell inequalities are strong witnesses in detecting multi-partite entanglement, since
different entanglement patterns lead to different quantum bounds\cite{Uffink2002,Nagata2002,Yusixia2003}.
We find that some quadratic Bell inequalities can be induced from the uncertainty relations of different pseudo observables on a single logical qubit.
Here we only give a simplest example to show that. First, let us focus on a useful theorem.

{\it Theorem.  --- } Let $\tilde{X}$ and $\tilde{Z}$ be two pseudo Pauli operators defined in Eq.(\ref{logical-Pauli-XYZ}). Consider two experimental Bell operators $\mB_{\vec{n}_1}^{\text{e}}$ and $\mB_{\vec{n}_2}^{\text{e}}$,  where $\vec{n}_i=(\sin\theta_i,\cos\theta_i)$ is a unit vector in the $xz$ plane ($i=1,2$), and $\mB_{\vec{n}_i}^{\text{e}}=2\sqrt{2}(\sin\theta_i\tilde{X}+\cos\theta_i\tilde{Z})$. Then
$\mB_{\vec{n}_1}^{\text{e}}$ and $\mB_{\vec{n}_2}^{\text{e}}$ satisfy the following uncertainty relation:
\begin{align}\label{uncertainty-Bell-operator}
\frac{(\langle\mB_{\vec{n}_1}^{\text{e}}\rangle+\langle\mB_{\vec{n}_2}^{\text{e}}\rangle)^2}{|\vec{n}_1+\vec{n}_2|^2}
+\frac{(\langle\mB_{\vec{n}_1}^{\text{e}}\rangle
-\langle\mB_{\vec{n}_2}^{\text{e}}\rangle)^2}{|\vec{n}_1-\vec{n}_2|^2}\leq8.
\end{align}

{\it Proof.} --- The detailed proof is shown in Appendix F. Similar uncertainty relations can be derived in other scenarios with more qubits.  \hfill $\blacksquare$

\begin{figure}[h]
\includegraphics[width=1\linewidth]{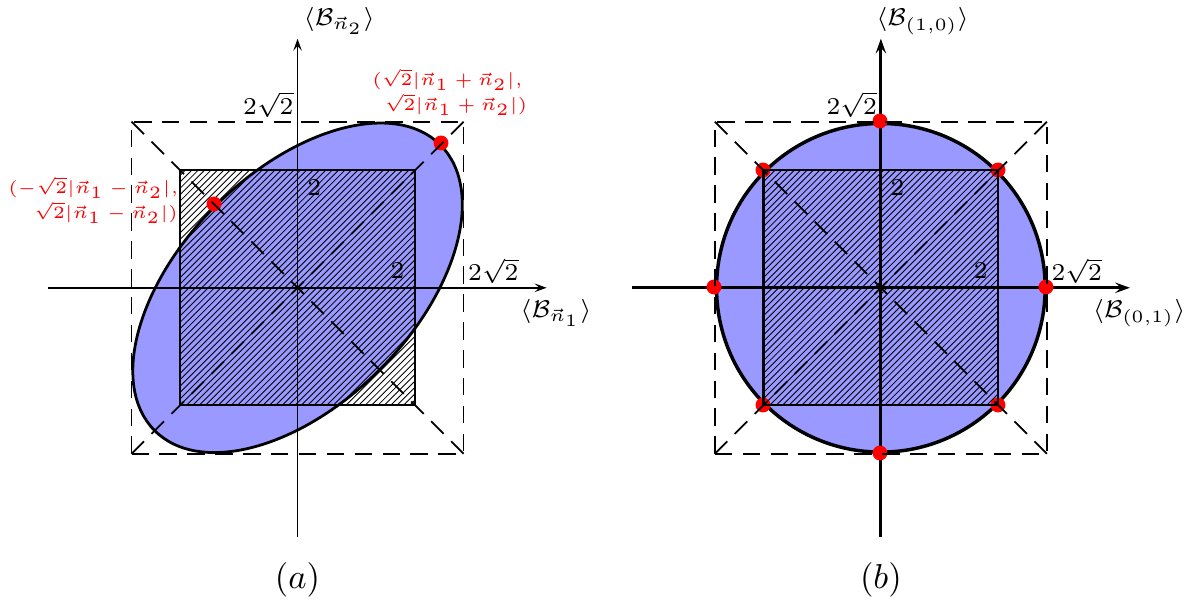}
\caption{(a) The uncertainty relation for two common of Bell operators in the two-qubit scenario. The theorem tells us that whatever the
given state is entangled or separable, the quantum expectations (even the classical expectations) for the two Bell operators should stay in the
blue  elliptic region.  Besides, from a linear Bell operator perspective, the range of their classical expectations is specified by the shadowed square.
 Namely, the classical expectations can only stay in the intersection of the shadowed square and the blue elliptic region.
 (b) A special example with $\vec{n}_1\perp\vec{n}_2$.
 The quantum expectations on the circle, $\langle\mB_{\vec{n}_1}^{\text{e}}\rangle^2+\langle\mB_{\vec{n}_2}^{\text{e}}\rangle^2=8$,
can only be attained by a maximally entangled state. Besides, note that the blue disc can cover all the
classical expectations.}\label{Bell-bound}
\end{figure}

Fig.\ref{Bell-bound}-(a) gives a geometric description of the above theorem. However, for simplicity,
we prefer to use the special uncertainty relation shown in Fig.\ref{Bell-bound}-(b),
i.e., $\langle\mB_{\vec{n}_1}^{\text{e}}\rangle^2+\langle\mB_{\vec{n}_2}^{\text{e}}\rangle^2\leq8$, to construct a quadratic Bell inequality,
where $\mB_{\vec{n}_1}^{\text{e}}=2\sqrt{2}\tilde{Z}$ and $\mB_{\vec{n}_1}^{\text{e}}=2\sqrt{2}\tilde{X}$.
Precisely, the construction
can be realized by replacing normal Pauli operators $Z_i,X_i$ in the expressions of
$\langle\mB_{\vec{n}_1}^{\text{e}}\rangle^2+\langle\mB_{\vec{n}_2}^{\text{e}}\rangle^2\leq8$ with
 $Z_i=\cos\phi_i A_i+\sin\phi_i B_i, X_i=-\sin\phi_i A_i+\cos\phi_i B_i$ respectively,
where $-\pi<\phi_i\leq\pi$.
The validation of such a replacement relies on the fact that
the disc for describing the  quantum expectations of $\mB_{\vec{n}_1}^{\text{e}}$ and $\mB_{\vec{n}_2}^{\text{e}}$ can cover
the (classical) square $-2\leq\langle\mB_{\vec{n}_1}^{\text{e}}\rangle_c,\langle\mB_{\vec{n}_1}^{\text{e}}\rangle_c\leq2$ completely
(see Figure.\ref{Bell-bound}-(b)).
Moreover, the total number of the measurement settings in this construction is less than that in the scheme based on Figure.\ref{Bell-bound}-(a).

{\it Example 4.} --- The two-qubit  Uffink's quadratic Bell inequality\cite{Uffink2002}. One can choose
$\mB_{\vec{n}_1}^{\text{e}}=2\sqrt{2}\tilde{Z}$ and $\mB_{\vec{n}_2}^{\text{e}}=2\sqrt{2}\tilde{X}$ to construct the inequality, and
replace $Z_1,X_1,Z_2,X_2$ in $\langle\mB_{\vec{n}_1}^{\text{e}}\rangle^2+\langle\mB_{\vec{n}_2}^{\text{e}}\rangle^2\leq8$
with $A_1, B_1,B_2,A_2$ respectively. Then the simplest Uffink's quadratic Bell inequality can be derived, i.e.,
\begin{align}\label{Uffink-2-qubit-quadratic-Bell-Inequality}
\langle A_1B_2+B_1A_2\rangle^2+\langle A_1A_2-B_1B_2\rangle^2\leq4.
\end{align}
Likewise, choosing $\mB_{\vec{n}_1}^{\text{e}}=2(\tilde{X}+\tilde{Z}),\mB_{\vec{n}_2}^{\text{e}}=2(\tilde{X}-\tilde{Z})$ and applying
the same replacement\cite{anotherreplacement} will give rise to the
Nagata-Koashi-Imoto quadratic Bell inequality\cite{Nagata2002}:
$\langle A_1(A_2+B_2)+B_1(A_2-B_2)\rangle^2+\langle A_1(A_2-B_2)-B_1(A_2+B_2)\rangle^2\leq8$.
\hfill$\blacksquare$

In summary, we find that the Bell operators chosen in most reported Bell tests have novel physical explanations, i.e.,  each of them
can be considered as a component of some pseudo spin (up to a constant).
This correspondence can not only provide an explicit physical significance for some kind of quantum violation (such as the maximal one)
of many famous Bell inequalities, but also inspire us to develop a new quantum-to-classical approach for the construction of Bell inequalities.
Compared with various previous quantum-to-classical approaches\cite{Lim2010,Salavrakos2017,Baccari2020,Cabello2008-1,Tang2013,Tang2017-1},
our approach applies to more kinds of Bell inequalities, including not only many
well-known Bell inequalities such as Mermin-type inequalities, CHSH- or Svetlichny-type inequalities
and the chained inequalities, but also some unnoticed graphical Bell inequalities.
Some of these Bell inequalities might have applications in some device-independent quantum information tasks,
e.g. self-testing\cite{Baccari2020}.  Besides, notice that
when the pseudo Pauli operators are defined based on nonlocal systems, they can be regarded as nonlocal observables.
A theorem in this work tells us that some quadratic Bell inequalities can be induced from
the uncertainty relations of such nonlocal observables. Conversely, the uncertainty relations of such nonlocal observables
can also be simulated by the quadratic Bell inequalities. This connection might open a new avenue for the study of uncertainty relations of some specific
nonlocal observables. Moreover,  whether composite logical qubits can induce useful Bell inequalities requires further investigation.

\begin{acknowledgements}
W. T. thanks K. Han, W. Du and D. Zhou for helpful discussions.
\end{acknowledgements}

\clearpage

\begin{appendices}

\section{Supplementary Materials}
\subsection{Appendix A: ~More choices for the logical basis}
Consider the scenarios that different pairs of Bell states are chosen as the bases. The corresponding
pseudo Pauli operators and the pseudo identity operators are shown  in Table.\ref{Table-logical-basis-Pauli-operator}.
\begin{center}
\begin{table*}
\caption{Pseudo Pauli operators, and the pseudo identity operator under different bases.} \label{Table-logical-basis-Pauli-operator}
\begin{center}
\begin{tabular}{c|c|c|c|c|c}
  \hline
   \hline
  {$|\tilde{0}\rangle$} & {$|\tilde{1}\rangle$} & {$\tilde{X}$} & {$\tilde{Z}$} & {$\tilde{Y}$} & {$\tilde{I}$}\\
   \hline
  $\frac{|00\rangle-|11\rangle}{\sqrt{2}}$ & $\frac{|01\rangle+|10\rangle}{\sqrt{2}}$ & $\frac{X\otimes Z+Z\otimes X}{2}$ & $\frac{Z\otimes Z-X\otimes X}{2}$ & $\frac{Y\otimes I+I\otimes Y}{2}$ & $\frac{I\otimes I+Y\otimes Y}{2}$ \\
  $\frac{|00\rangle-|11\rangle}{\sqrt{2}}$ & $\frac{|01\rangle-|10\rangle}{\sqrt{2}}$ & $\frac{I\otimes X-X\otimes I}{2}$ & $\frac{Z\otimes Z+Y\otimes Y}{2}$ & $\frac{Z\otimes Y-Y\otimes Z}{2}$ & $\frac{I\otimes I-X\otimes X}{2}$ \\
  $\frac{|00\rangle-|11\rangle}{\sqrt{2}}$ & $\frac{|00\rangle+|11\rangle}{\sqrt{2}}$ & $\frac{I\otimes Z+Z\otimes I}{2}$ & $\frac{Y\otimes Y-X\otimes X}{2}$ & $\frac{Y\otimes X-X\otimes Y}{2}$ & $\frac{I\otimes I+Z\otimes Z}{2}$ \\
  $\frac{|01\rangle+|10\rangle}{\sqrt{2}}$ & $\frac{|01\rangle-|10\rangle}{\sqrt{2}}$ & $\frac{Z\otimes I-I\otimes Z}{2}$ & $\frac{X\otimes X+Y\otimes Y}{2}$ & $\frac{X\otimes Y-Y\otimes X}{2}$ & $\frac{I\otimes I-Z\otimes Z}{2}$ \\
  $\frac{|00\rangle+|11\rangle}{\sqrt{2}}$ & $\frac{|01\rangle+|10\rangle}{\sqrt{2}}$ &  $\frac{I\otimes X+X\otimes I}{2}$ & $\frac{Z\otimes Z-Y\otimes Y}{2}$ & $\frac{Z\otimes Y+Y\otimes Z}{2}$ & $\frac{I\otimes I+X\otimes X}{2}$ \\
   $\frac{|00\rangle+|11\rangle}{\sqrt{2}}$ & $\frac{|01\rangle-|10\rangle}{\sqrt{2}}$  & $\frac{Z\otimes X-X\otimes Z}{2}$ & $\frac{Z\otimes Z+X\otimes X}{2}$ & $\frac{I\otimes Y-Y\otimes I}{2}$ & $\frac{I\otimes I-Y\otimes Y}{2}$ \\
  \hline
   \hline
\end{tabular}
\end{center}
\end{table*}
\end{center}
Apart from Bell states, the basis can also be chosen from other maximally entangled states. To give an example,  first we rewrite the logical qubit as follows,
\begin{align}\label{logical-qubit}
    |\tilde{\psi}\rangle=\cos\frac{\theta}{2}|\tilde{0}\rangle+e^{i\varphi}\sin\frac{\theta}{2}|\tilde{1}\rangle,
\end{align}
where $\theta\in[0,\pi]$ and $\varphi\in[0,2\pi)$. In fact, $|\tilde{\psi}\rangle$ can be considered as an eigenstate of
$\vec{a}\cdot\vec{\tilde{\sigma}}=a^x\tilde{X}+a^y\tilde{Y}+a^z\tilde{Z}$, where $\vec{a}=(a^x,a^y,a^z)$ is a real vector,
 and $\vec{\tilde{\sigma}}=(\tilde{X},\tilde{Y},\tilde{Z})$. Moreover, one can simply take $\varphi=0$ (or $\pi$) if $n_y$ vanishes,
and get the following observation.

{\it Observation 1.}--- The eigenstate of $\tilde{Z}(\theta):=\tilde{Z}\cos\theta+\tilde{X}\sin\theta$ associated to the eigenvalue $+1$ can be written as
\begin{align}\label{maximal-entangled-XZ}
|\tilde{0}(\theta)\rangle=\cos\frac{\theta}{2}|\tilde{0}\rangle+\sin\frac{\theta}{2}|\tilde{1}\rangle
=\frac{1}{\sqrt{2}}[|0\rangle|0(\theta)\rangle+|1\rangle|1(\theta)\rangle],
\end{align}
indicating that it is also a maximally entangled state which can be stabilized by $Z_1Z_2(\theta)$ and $X_1X_2(\theta)$, where
$Z(\theta)=Z\cos\theta+X\sin\theta$, $X(\theta)=-Z\sin\theta+X\cos\theta$,
 $|0(\theta)\rangle=\cos\frac{\theta}{2}|0\rangle+\sin\frac{\theta}{2}|1\rangle$ and $|1(\theta)\rangle=-\sin\frac{\theta}{2}|0\rangle+\cos\frac{\theta}{2}|1\rangle$.

Notice that $Z(\theta)$ and $X(\theta)$ are mutually complementary (most noncommutative) observables (similar to $X$ and $Z$),
and besides, $|0(\theta)\rangle$ and $|1(\theta)\rangle$ are two eigenstates of $Z(\theta)$. Analogous to Eq.(\ref{CHSH-op-stabilizer-rep}) in the main text,
one can construct a more general Bell operator as follows,
\begin{align}\label{CHSH-op-XZ-plane-L}
   \mB_{\text{CHSH}}^{\text{e}}(L)=2\sqrt{2}\tilde{Z}(\theta),
\end{align}
and accordingly,
\begin{align}\label{CHSH-op-XZ-plane-S}
&\mB_{\text{CHSH}}^{\text{e}}(S)=\sqrt{2}(X_1 X_2(\theta)+Z_1 Z_2(\theta))\cr
   =&\sqrt{2}[X_1(X_2\cos\theta-Z_2\sin\theta)+Z_1(Z_2\cos\theta+X_2\sin\theta)].\cr
\end{align}
In real experiment, we should decomposed $X_1X_2(\theta)$ and $Z_1Z_2(\theta)$ into
$X_1\frac{X_2(\theta)+Z_2(\theta)}{\sqrt{2}}+X_1 \frac{X_2(\theta)-Z_2(\theta)}{\sqrt{2}}$ and
$Z_1 \frac{X_2(\theta)+Z_2(\theta)}{\sqrt{2}}-Z_1\frac{X_2(\theta)-Z_2(\theta)}{\sqrt{2}}$, respectively.
Accordingly, the associated Bell operators is
\begin{align}\label{CHSH-op-XZ-plane-realexp}
   \mB_{\text{CHSH}}^{\text{e}}
   =&X_1[X_2\sin(\theta+\frac{\pi}{4})+Z_2\cos(\theta+\frac{\pi}{4})]\cr
   &+X_1[X_2\cos(\theta+\frac{\pi}{4})-Z_2\sin(\theta+\frac{\pi}{4})]\cr
   &+Z_1[X_2\sin(\theta+\frac{\pi}{4})+Z_2\cos(\theta+\frac{\pi}{4})]\cr
   &-Z_1[X_2\cos(\theta+\frac{\pi}{4})-Z_2\sin(\theta+\frac{\pi}{4})].\cr
\end{align}
By replacing $X_1,Z_1,X_2\sin(\theta+\frac{\pi}{4})+Z_2\cos(\theta+\frac{\pi}{4})$
and $X_2\cos(\theta+\frac{\pi}{4})-Z_2\sin(\theta+\frac{\pi}{4})$ with $A_1,A_1^{\prime},B_2$ and $B_2^{\prime}$,
one can derive the final Bell inequality.

\subsection{Appendix B: ~Using sum of square technique prove maximal quantum violations of some Bell inequalities}

As mentioned in the main text, we prefer to choose two mutually orthogonal and fully entangled stabilizer states as the basis. Besides, it is noted
that each stabilizer state is locally unitary equivalent to the graph state. Then the logical qubit basis can be chosen from graph state basis.
Precisely, choose $|\tilde{0}\rangle=|G_n\rangle$ and $|\tilde{1}\rangle=Z_C|G_n\rangle$, where $|G_n\rangle$ is the $n$-qubit graph state
stabilized by $g_i$ ($i=1,2,\cdots,n$), and $Z_C\equiv\otimes_{i\in C}Z_i$ ($C\subset\{1,2,\cdots,n\}$). Notice that
$|\tilde{0}\rangle\langle\tilde{0}|=|G_n\rangle\langle G_n|=\frac{1}{2^n}\prod_{i=1}^n(I+g_i)$ and
$|\tilde{1}\rangle\langle\tilde{1}|=Z_C|G_n\rangle\langle G_n|Z_C=\frac{1}{2^n}Z_C(\prod_{i=1}^n(I+g_i))Z_C$. Thus,
\begin{align}\label{graph-Pse-Pauli}
    \tilde{Z}=&\frac{1}{2^n}[\prod_{i=1}^n(I+g_i)-Z_C(\prod_{i=1}^n(I+g_i))Z_C],\cr
    \tilde{X}=&\frac{1}{2^n}[\prod_{i=1}^n(I+g_i)+Z_C(\prod_{i=1}^n(I+g_i))Z_C]Z_C.
\end{align}
 Assume that there are $m$
independent stabilizers anti-commutate with $Z_C$, i.e., $n-m$ independent stabilizers commutate with $Z_C$. In the group generated by
$g_1,g_2,\cdots,g_n$, there are a total of $2^{n-m}(C_m^0+C_m^2+C_m^4+\cdots)=2^{n-m}2^{m-1}=2^{n-1}$ terms commutating with $Z_C$ (each term
contains a product of an even number of (independent) stabilizers anti-commutating with $Z_C$).
Namely,  the expansion of $\prod_{i=1}^n(I+g_i)$ has $2^n$ terms, but only half of them ($2^{n-1}$) commutate with $Z_C$.
Therefore, both $\tilde{Z}$ and $\tilde{X}$ have $2^{n-1}$ terms.

First, consider the scenario for the construction of Mermin-type Bell inequality. One can choose $\mB^e(L)=2^{n-1}\tilde{Z}$.
If the Bell inequality can be constructed successfully, i.e., by replacing $X_i$ and $Z_i$ in $\tilde{Z}$ with $A_i$ and $B_i$. The maximal quantum
violation is $2^{n-1}$ (which equals to the total number of terms), one can get that directly from the fact that each term in $\mB^e(L)$ is a dichotomic operator (Pauli observable).

Next, we consider the scenario for the construction of Svetlichny-type (or CHSH-type) Bell inequality.
Take the choice of $\mB^e(L)=2^{n-1}\sqrt{2}\frac{\tilde{Z}+\tilde{X}}{\sqrt{2}}$ as an example.
Note that there are $2^n$ terms in the $\mB^e(S)=\mB^e$. One can also denote
$2^{n-1}\tilde{Z}=\sum_{k=1}^{2^{n-1}}P_k(X_1,\cdots,X_n,Z_1,\cdots,Z_n)$ and
$2^{n-1}\tilde{X}=\sum_{l=1}^{2^{n-1}}Q_l(X_1,\cdots,X_n,Z_1,\cdots,Z_n)$, i.e.,
 $P_k(X_1,\cdots,X_n,Z_1,\cdots,Z_n)$ is a term from the expansion of $2^{n-1}\tilde{Z}$, and
 $Q_j(X_1,\cdots,X_n,Z_1,\cdots,Z_n)$ is a term from the expansion of $2^{n-1}\tilde{X}$.
After replacing
$X_i$ and $Z_i$ in the expansion of $\mB^e$  with $A_i$ and $B_i$, one can get the final Bell operator
$\mB=\sum_{k=1}^{2^{n-1}}P_k(A_1,\cdots,A_n,B_1,\cdots,B_n)+\sum_{l=k}^{2^{n-1}}Q_k(A_1,\cdots,A_n,B_1,\cdots,B_n)$
(also has $2^n$ terms in its expression). For simplicity, sometimes we also write
$\mB=\sum_{k=1}^{2^{n-1}}P_k+\sum_{k=1}^{2^{n-1}}Q_k$.

Consider the following sum of square,
\begin{align}\label{SOS}
    2^{n-1}\sqrt{2} I-\mB=\frac{1}{\sqrt{2}}\sum_{k=1}^{2^{n-1}}(I-\frac{P_k+Q_k}{\sqrt{2}})^2.
\end{align}
It holds true if $\sum_{k=1}^{2^{n-1}}\{P_k,Q_k\}=0$, where $\{P_k,Q_k\}=P_kQ_k+Q_kP_k$. Precisely, if there exists a partition
$\{P_k|k=1,2,\cdots,2^{n-1}\}\cup\{Q_k|k=1,2,\cdots,2^{n-1}\}=\bigcup_{r\in\mS_1,s\in\mS_2}\{P_r,Q_r,P_s,Q_s\}$ in which
 $\{P_r,Q_r\}=-\{P_s,Q_s\}$, then Eq.(\ref{SOS}) holds,
where $\mS_1\bigcup\mS_2=\{1,2,\cdots,2^{n-1}\}$ and $|\mS_1|=|\mS_2|=2^{n-2}$. This condition can be satisfied only if $\tilde{X}$
and $\tilde{Z}$ have some particular forms, i.e.,  the bases
$|\tilde{0}\rangle$ and $|\tilde{0}\rangle$ should be specially chosen,
 such as Bell states, GHZ states, and several specific families of graph states.
In these scenarios, Eq.(\ref{SOS}) implies that $2^{n-1}\sqrt{2} I-\mB\succeq0$. Also notice that in $P_k$ and $Q_k$ from Eq.(\ref{SOS}),
$A_i$ and $B_i$ are normal local dichotomic observables, then $2^{n-1}\sqrt{2}$ is the maximal quantum violation
of an $n$-qubit Bell inequality. This maximal quantum violation can be realized (Eq.(\ref{SOS}) turns to an equality) by choosing $A_i=X_i$ and $B_i=Z_i$.

For example, choose
$|\tilde{0}\rangle=(|00\rangle+|11\rangle)/\sqrt{2}$ and $|\tilde{1}\rangle=(|01\rangle-|10\rangle)/\sqrt{2}$, one can get
 $\mB=A_1B_2+A_1^{\prime}B_2+A_1B_2^{\prime}-A_1^{\prime}B_2^{\prime}$. Let $P_1=A_1B_2,Q_1=A_1B_2^{\prime}$,
$P_2=A_1^{\prime}B_2,Q_2=-A_1^{\prime}B_2^{\prime}$. Clearly $\{P_1,Q_1\}=-\{P_2,Q_2\}$. The sum of square
$2\sqrt{2}I-\mB$ can be represented as $\frac{1}{\sqrt{2}}[(I-\frac{A_1B_2+A_1B_2^{\prime}}{\sqrt{2}})^2
+(I-\frac{A_1^{\prime}B_2-A_1^{\prime}B_2^{\prime}}{\sqrt{2}})^2]$. Then the maximal quantum violation
of this CHSH inequality is $2\sqrt{2}$.

Apart from Eq.(\ref{SOS}), there are many other decompositions, and some decompositions may involve higher orders of
 $\langle\mB\rangle_{\max}I-\mB$ (e.g
$(\langle\mB\rangle_{\max}I-\mB)^2$). How to realize that for a general $\langle\mB\rangle_{\max}I-\mB$ remains an open question.
As far as I know, the  decompositions for Svetlichny-type (or CHSH-type),
chained, and some special graphical Bell inequalities have been studied and many useful results have been reported in previous literature.

In this work, the exact maximal quantum violations for the Mermin-type Bell inequalities can be derived from the expression of $\mB^{e}(L)$
directly (each term of the corresponding $\mB$ is a dichotomic observable).
For the  Svetlichny-type (or CHSH-type), and chained Bell inequalities,  based  on the sum of square method,
we can give their exact maximal quantum violations. However, for the three Bell inequalities given in Example 2,
we  give the exact maximal quantum violation for the first one, while for the other two,
we only give minimum estimations for their maximal quantum violations
(Despite that, it can still grantee the corresponding gap between this rough upper quantum bound and the classical one large enough for the demonstration of nonlocality).

\subsection{Appendix C: ~More detailed description for constructing three typical Bell inequalities}

Table \ref{TB2} shows us one of the constructions for each Bell inequality. For the CHSH inequality and the Svetlichny, we have listed two
choices for $\mB^{e}(L)$. Below we show the construction of these two Bell inequalities under the other choices.

For the  CHSH inequality, apart from choosing $2\sqrt{2}\tilde{Z}$ for $\mB^{e}(L)$ listed in Table \ref{TB2},
one can also choose $\mB^{e}(L)=2\sqrt{2}\frac{\tilde{X}+\tilde{Z}}{\sqrt{2}}$.
Accordingly, $\mB^{e}=\mB^{e}(S)=Z_1X_2-X_1Z_2+X_1X_2+Z_1Z_2$. By using the replacements $A_1=Z_1,A_1^{\prime}=X_1,B_2=X_2,B_2^{\prime}=Z_2$, one can get
$\mB=A_1 B_2+A_1B_2^{\prime}+A_1^{\prime}B_2-A_1^{\prime}B_2^{\prime}$. Based on the classical inequality $|a+b+c-abc|\leq2$ ($|a|,|b|,|c|\in[-1,1]$),
one can get $|\mB|\leq2$, while quantum upper bound of $\mB$ is $2\sqrt{2}$.

For the  Svetlichny inequality, we consider the choice  $\mB^{e}(L)=4\sqrt{2}\tilde{Z}$,
where the corresponding $\tilde{Z}$ has been listed in Table \ref{TB2}.
Then $\mB^{e}(S)=\sqrt{2}(Z_1Z_2Z_3-Z_1X_2X_3-X_1Z_2X_3-X_1X_2Z_3)$, and
$\mB^{e}=Z_1Z_2(\frac{X_3+Z_3}{\sqrt{2}}-\frac{X_3-Z_3}{\sqrt{2}})-Z_1X_2(\frac{X_3+Z_3}{\sqrt{2}}+\frac{X_3-Z_3}{\sqrt{2}})
-X_1Z_2(\frac{X_3+Z_3}{\sqrt{2}}+\frac{X_3-Z_3}{\sqrt{2}})-X_1X_2(\frac{X_3+Z_3}{\sqrt{2}}-\frac{X_3-Z_3}{\sqrt{2}})$. The replacements we adopt here
are $A_{1}=Z_{1},A_{2}=Z_{2},A_3=-\frac{X_3+Z_3}{\sqrt{2}},B_{1}=X_{1},B_{2}=X_{2},B_3=-\frac{X_3-Z_3}{\sqrt{2}}$, giving rise to
$\mB=A_1B_2B_3+B_1A_2B_3+B_1B_2A_3-A_1A_2A_3+B_1A_2A_3+A_1B_2A_3+A_1A_2B_3-B_1B_2B_3$. Using the sum of square technique, one can prove
the quantum upper bound of $\mB$ is $4\sqrt{2}$. Since the expression of $\mB$ contains two
Mermin inequalities, and the classical bound for each Mermin inequality is $2$, therefore,
one have $|\mB|\leq4$.

\begin{table*}\caption{Detailed constructions for three typical Bell inequalities.} \label{TB2}
    \centering
    \begin{tabular}{c|ccc}
    \hline
    \hline
        & CHSH  & Mermin & Svetlichny   \\
  \hline
     \parbox[c]{2.2cm}{$|\tilde{0}\rangle$ }  & $\frac{1}{\sqrt{2}}(|00\rangle+|11\rangle)$ &  $\frac{1}{2}[|0\rangle(|00\rangle-|11\rangle)-|1\rangle(|01\rangle+|10\rangle)]$ &  $\frac{1}{2}[|0\rangle(|00\rangle-|11\rangle)-|1\rangle(|01\rangle+|10\rangle)]$  \\
     \parbox[c]{2.2cm}{$|\tilde{1}\rangle$ }  & $\frac{1}{\sqrt{2}}(|01\rangle-|10\rangle)$ &
     $\frac{1}{2}[|0\rangle(|01\rangle+|10\rangle)+|1\rangle(|00\rangle-|11\rangle)]$ & $\frac{1}{2}[|0\rangle(|01\rangle+|10\rangle)+|1\rangle(|00\rangle-|11\rangle)]$ \\
     \parbox[c]{2.2cm}{$\tilde{X}$ }  & $\frac{1}{2}(Z_1X_2-X_1Z_2)$ & \thead{$\frac{1}{4}(X_1Z_2Z_3+Z_1X_2Z_3$ \\ $+Z_1Z_2X_3-X_1X_2X_3)$} & \thead{$\frac{1}{4}(X_1Z_2Z_3+Z_1X_2Z_3$ \\ $+Z_1Z_2X_3-X_1X_2X_3)$} \\
     \parbox[c]{2.2cm}{$\tilde{Z}$ }  & $\frac{1}{2}(X_1X_2+Z_1Z_2)$ & \thead{$\frac{1}{4}(Z_1Z_2Z_3-Z_1X_2X_3$ \\ $-X_1Z_2X_3-X_1X_2Z_3)$} & \thead{$\frac{1}{4}(Z_1Z_2Z_3-Z_1X_2X_3$ \\ $-X_1Z_2X_3-X_1X_2Z_3)$} \\
      \parbox[c]{2.2cm}{ $\mB^{e}(L)$ }  &  $2\sqrt{2}\tilde{Z}$ & $4\tilde{Z}$ & $4\sqrt{2}\frac{\tilde{X}-\tilde{Z}}{\sqrt{2}}$ \\
      \parbox[c]{2.2cm}{ $\langle\mB^e\rangle_{\max}$ }  & $2\sqrt{2}$ & $4$ & $4\sqrt{2}$\\
           \parbox[c]{2.2cm}{$|\psi\rangle_{\max}$} & \thead{$\frac{1}{\sqrt{2}}(|00\rangle+|11\rangle)$} & \thead{$\frac{1}{\sqrt{2}}(|\circlearrowleft\circlearrowleft\circlearrowleft\rangle+|\circlearrowright\circlearrowright\circlearrowright\rangle)$}
    &\thead{$\frac{1}{\sqrt{2}}[(e^{-i\frac{3\pi}{8}}|0\rangle+e^{i\frac{\pi}{8}}|1\rangle)|\circlearrowleft\circlearrowleft\rangle$ \\
    $+(e^{i\frac{3\pi}{8}}|0\rangle+e^{-i\frac{\pi}{8}}|1\rangle)|\circlearrowright\circlearrowright\rangle]$} \\
      \parbox[c]{2.2cm}{ $\mB^{e}(S)$ }  & $\sqrt{2}(X_1X_2+Z_1Z_2)$ & \thead{$Z_1Z_2Z_3-Z_1X_2X_3$ \\ $-X_1Z_2X_3-X_1X_2Z_3$} & \thead{ $\sqrt{2}[\frac{X_1+Z_1}{\sqrt{2}}X_2Z_3+\frac{Z_1+X_1}{\sqrt{2}}Z_2X_3$ \\ $+\frac{X_1-Z_1}{\sqrt{2}}Z_2Z_3-\frac{X_1-Z_1}{\sqrt{2}}X_2X_3$ ]}\\
        \parbox[c]{2.2cm}{ $\mB^e$ } & \thead{$X_1\frac{X_2+Z_2}{\sqrt{2}}+ X_1\frac{X_2-Z_2}{\sqrt{2}}$ \\
    $+Z_1\frac{X_2+Z_2}{\sqrt{2}}- Z_1\frac{X_2-Z_2}{\sqrt{2}}$}          & \thead{$Z_1Z_2Z_3-Z_1X_2X_3$ \\ $-X_1Z_2X_3-X_1X_2Z_3$}  & \thead{$X_1Z_2Z_3+Z_1X_2Z_3+Z_1Z_2X_3-X_1X_2X_3$ \\ $+Z_1X_2X_3+X_1Z_2X_3+X_1X_2Z_3-Z_1Z_2Z_3$} \\
      \parbox[c]{2.2cm}{ $\mB$} & \thead{$A_1 B_2+A_1B_2^{\prime}$\\ $+A_1^{\prime}B_2-A_1^{\prime}B_2^{\prime}$}         & \thead{$A_1A_2A_3-A_1B_2B_3$\\ $-B_1A_2B_3-B_1B_2A_3$}  & \thead{$A_1B_2B_3+B_1A_2B_3+B_1B_2A_3-A_1A_2A_3$\\ $+B_1A_2A_3+A_1B_2A_3+A_1A_2B_3-B_1B_2B_3$} \\
         \parbox[c]{2.2cm}{ Bell inequality }  & $\langle\mB\rangle_c\leq2$ & $\langle\mB\rangle_c\leq2$ & $\langle\mB\rangle_c\leq4$\\
     \hline
     \hline
    \end{tabular}
\end{table*}

\subsection{Appendix D: ~The calculation for pseudo Pauli operators in the $[[5,1,3]]$ coding space}

It is noticed that two bases in the $[[5,1,3]]$ coding space are $|L_{5}\rangle$ and $Z_0Z_1Z_2Z_3Z_4|L_{5}\rangle$, and
$|L_{5}\rangle\langle L_{5}|=\prod_{i=0}^{4}\frac{I+g_i}{2}$, where $g_{i\oplus1}=Z_{i}X_{i\oplus1}Z_{i\oplus2}$ are the $i$-th stabilizer of $|L_{5}\rangle$, where $i\in\{0,1,2,3,4\}$.

Since we have chosen
\begin{align*}
    |\tilde{0}\rangle=|L_5\rangle,~~|\tilde{1}\rangle=Z_0Z_1Z_2Z_3Z_4|L_5\rangle.
\end{align*}
in the maintext, the logical Pauli operators $\tilde{Z}$, $\tilde{X}$ and $\tilde{Y}$ can be calculated based on the following relations:
\begin{align*}
    \tilde{Z}=&|\tilde{0}\rangle\langle\tilde{0}|-|\tilde{1}\rangle\langle\tilde{1}|,\cr
    \tilde{X}=&|\tilde{0}\rangle\langle\tilde{1}|+|\tilde{1}\rangle\langle\tilde{0}|,\cr
    \tilde{Y}=&i(|\tilde{1}\rangle\langle\tilde{0}|-|\tilde{0}\rangle\langle\tilde{1}|),\cr
    \tilde{I}=&|\tilde{0}\rangle\langle\tilde{0}|+|\tilde{1}\rangle\langle\tilde{1}|.\cr
\end{align*}

Therefore,
\begin{align*}
        \tilde{Z}=&\frac{1}{2^5}[\prod_{i=0}^{4}(I+g_i)\cr
        &-Z_0Z_1Z_2Z_3Z_4\prod_{i=0}^{4}(I+g_i)Z_0Z_1Z_2Z_3Z_4]\cr
        =&\frac{1}{16}(\sum_{i=0}^4g_i+\sum_{i\neq j\neq k=0}^4g_ig_jg_k+g_0g_1g_2g_3g_4)\cr
    =&\frac{1}{16}[\sum_{i=0}^4(Z_{i}X_{i\oplus1}Z_{i\oplus2}-Z_iY_{i\oplus1}X_{i\oplus2}Y_{i\oplus3}Z_{i\oplus4}\cr
    &+X_iY_{i\oplus2}Y_{i\oplus3})-X_0X_1X_2X_3X_4].
\end{align*}

Likewise, one can get
\begin{align*}
    \tilde{X}
    =&\frac{Z_0Z_1Z_2Z_3Z_4}{16}(I+\sum_{i\neq j=0}^4g_ig_j\cr
    &+\sum_{i\neq j\neq k\neq l=0}^4g_ig_jg_kg_l)\cr
    =&\frac{1}{16}[\sum_{i=0}^4(-Z_iX_{i\oplus2}X_{i\oplus3}-Y_iZ_{i\oplus1}Y_{i\oplus2}\cr
    &+Y_iX_{i\oplus1}Z_{i\oplus2}X_{i\oplus3}Y_{i\oplus4})+Z_0Z_1Z_2Z_3Z_4],\cr
    \end{align*}
    \begin{align*}
    \tilde{Y}
    =&\frac{iZ_0Z_1Z_2Z_3Z_4}{16}(\sum_{i=0}^4g_i\cr
    &+\sum_{i\neq j\neq k=0}^4g_ig_jg_k+g_0g_1g_2g_3g_4)\cr
    =&\frac{1}{16}[\sum_{i=0}^4(-Y_iZ_{i\oplus2}Z_{i\oplus3}-X_iY_{i\oplus1}X_{i\oplus2}\cr
    &+X_iZ_{i\oplus1}Y_{i\oplus2}Z_{i\oplus3}X_{i\oplus4})+Y_0Y_1Y_2Y_3Y_4],
\end{align*}
and
    \begin{align*}
    \tilde{I}
    =&\frac{1}{2^5}[\prod_{i=0}^{4}(I+g_i)\cr
        &+Z_0Z_1Z_2Z_3Z_4\prod_{i=0}^{4}(I+g_i)Z_0Z_1Z_2Z_3Z_4]\cr
     =&\frac{1}{16}(I+\sum_{i\neq j=0}^4g_ig_j
    +\sum_{i\neq j\neq k\neq l=0}^4g_ig_jg_kg_l)\cr
    =&\frac{1}{16}[I+\sum_{i=0}^4(Y_iX_{i\oplus1}X_{i\oplus2}Y_{i\oplus3}-Z_iY_{i\oplus1}Y_{i\oplus2}Z_{i\oplus3}\cr
    &+X_iZ_{i\oplus1}Z_{i\oplus2}X_{i\oplus3})].
\end{align*}

Since the forms of $\tilde{Z},\tilde{X}$ and $\tilde{Y}$ have similar algebraic structures, they can induce similar Bell inequalities.
Besides, even based on $\tilde{I}$, one can construct a Bell inequality. To see that, we start from such a Bell operator
\begin{align}\label{APPDIX-Logical-Bell-I}
    \mB_C(L)=16\tilde{I}.
\end{align}
After a similar discussion in the main text, one can get the final Bell operator
\begin{align}\label{APPDIX-Bell-I-op}
    \mB_C=&I+\sum_{i=0}^4(C_iB_{i\oplus1}B_{i\oplus2}C_{i\oplus3}-A_iC_{i\oplus1}C_{i\oplus2}A_{i\oplus3}\cr
    &+B_iA_{i\oplus1}A_{i\oplus2}B_{i\oplus3}).
\end{align}
By a numerical calculation, one can get the Bell inequality as follows:
\begin{align}\label{APPDIX-Bell-I-ineq}
    -6\leq\langle\mB\rangle_c\leq10.
\end{align}
By contrast, its quantum maximal violation is at least 16 according to Eq.(\ref{APPDIX-Logical-Bell-I}). On the other hand, since each term
in Eq.(\ref{APPDIX-Bell-I-op}) is a dichotomic observable (except $I$), and there are a total of 16 terms, the quantum maximal violation of
Eq.(\ref{APPDIX-Bell-I-ineq}) is at most 16. Therefore, the quantum maximal violation can only be 16,
which can be attained by any state (including mixed states) in the
Hilbert space spanned by the bases $|L_5\rangle$ and $Z_1Z_2Z_3Z_4Z_5|L_5\rangle$. It follows that all the states in this space are highly entangled, and the nonlocality exhibited by these states are very strong.

\subsection{Appendix E: ~The Mermin inequality and the Svetlichny inequality for $n$-qubit ($n\geq3$)  scenario}

For the $n$-qubit ($n\geq3$) Mermin inequality, a typical family of the choices for  $\mB^e_n(L)$ can be described as $\{2^{n-1}(\cos\theta\tilde{Z}^{(n)}+\sin\theta\tilde{X}^{(n)})|\theta\in[0,\pi]\}$. Clearly, one have $\langle\mB^e_n(L)\rangle_{\max}=2^{n-1}$. For simplicity, we only consider the special case with $\theta=0$, i.e.,
$\mB^e_n(L)=2^{n-1}\tilde{Z}^{(n)}$ (another simple choice is $\mB^e_n(L)=2^{n-1}\tilde{X}^{(n)}$). Rewriting $\mB^e_n(L)$ in terms of a linear combination of stabilizers will give rise to $\mB^e_n(S)$, and one can check that
there are $2^{n-1}$ items in the corresponding $\mB^e_n(S)$. Then replacing each $X_i$ and $Z_i$ in $\mB^e_n(S)$ with $A_i$ and $B_i$ respectively, one can get the Bell operator $\mB_n$. Note that the algebraic structures of $\mB_n$ and $\mB^e_n(S)$ are the same. This indicates that one can also calculate the
classical upper bound  based on $\mB^e_n(S)$ (just for convenience, the following derivation also works if one replace
each $X_i$ and $Z_i$ with $A_i$ and $B_i$). In the following  we shall use the value assignment technique and some inequality techniques
to derive this bound.

Note that in the experiment, the outcome of measuring $X_i$ (or $Z_i$) on the $i$-th qubit is either $+1$ or $-1$.  Assume that the whole quantum system admits a LHV model, then one can assign values $\pm$ to all the Pauli operators in the expression of $\mB^e_n(S)$.
Here we should stress that value assignment only applies to physical qubit (normal)  Pauli operators rather than logical qubit ones. Namely, for each run of the value assignment, one should first rewrite $\tilde{Z}^{(n)}$ and $\tilde{X}^{(n)}$ in terms of normal Pauli operators (such as Eq.(\ref{logical-Pauli-XYZ}) in the main text), then calculate their values (they might be different in different runs).
For simplicity,  denote by $v(\tilde{Z}^{(n)})$ and $v(\tilde{X}^{(n)})$ the classical value of $\tilde{Z}^{(n)}$ and $\tilde{X}^{(n)}$ respectively.

 Notice that applying $\ast_1$ and $\ast_2$ sequentially on
$\tilde{Z}^{(1)}=Z$ (or $\tilde{X}^{(1)}=X$) will give rise to $\tilde{Z}^{(3)}$ (or $\tilde{X}^{(3)}$).  More generally, applying $\ast_n$ and $\ast_{n+1}$ sequentially on
$\tilde{Z}^{(n)}$, one can get $\tilde{Z}^{(n+2)}$. Likewise, $\tilde{X}^{(n+2)}$ can be derived by applying the same operations on $\tilde{X}^{(n)}$. This observation will play a key role in the following argument. Inspired by that, one can calculate the (classical) maximum of $v(\tilde{Z}^{(n+2)})$ by exploiting the maximum of $v(\tilde{Z}^{(n)})$, where $n\geq3$.

For the $3$-qubit scenario, notice that $\mB^{\text{e}}_3(S)=Z_1Z_2Z_3+Z_1X_2X_3+X_1Z_2X_3-X_1X_2Z_3$ (which is locally unitary equivalent to the standard form in Table {\ref{TB2}}). Clearly, one can get $\langle\mB_3\rangle_c=\langle\mB^{\text{e}}_3(S)\rangle_c\leq2$. Accordingly, one have $v(\tilde{Z}^{(3)})\leq2/4=1/2$ (by contrast, the maximal quantum value of $\tilde{Z}^{(3)}$ is $+1$ (the maxinal eigenvalue)). Similarly, one can also check that $v(\tilde{X}^{(3)})\leq2/4=1/2$.

For the $4$-qubit scenario, we can invoke Eq.(\ref{logical-Pauli-XYZ}) in the main text, i.e.,
$\tilde{Z}^{(2)}=(Z_1Z_2+X_1X_2)/2$ and $\tilde{X}^{(2)}=(Z_1X_2-X_1Z_2)/2$. Applying $\ast_2$ and $\ast_3$ sequentially on
$\tilde{Z}^{(2)}$, one can get $\tilde{Z}^{(4)}$ as follows,
\begin{align}\label{Z-4-Z-2-rep}
    \tilde{Z}^{(4)}=\frac{1}{2}(Z_1\tilde{Z}^{(3)}_{\{2,3,4\}}+X_1\tilde{X}^{(3)}_{\{2,3,4\}}).
\end{align}
Here,  $\tilde{Z}^{(3)}_{\{2,3,4\}}$ and $\tilde{X}^{(3)}_{\{2,3,4\}}$ represent $\tilde{Z}^{(3)}$ and $\tilde{X}^{(3)}$ defined on the subsystem $\{2,3,4\}\subset\{1,2,3,4\}$.
Likewise, $\ast_2$ and $\ast_3$ acting on $\tilde{X}^{(2)}$ will induce
\begin{align}\label{X-4-X-2-rep}
    \tilde{X}^{(4)}=\frac{1}{2}(Z_1\tilde{X}^{(3)}_{\{2,3,4\}}-X_1\tilde{Z}^{(3)}_{\{2,3,4\}}).
\end{align}

Since the value assignments for $X_1$ and $Z_1$ satisfy $v(X_1),v(Z_1)\in\{+1,-1\}$,  one have
\begin{align*}
    v(\tilde{Z}^{(4)})\leq &|v(\tilde{Z}^{(4)})|\cr
\leq& \frac{1}{2}|v(Z_1)v(\tilde{Z}^{(3)}_{\{2,3,4\}})+v(X_1)v(\tilde{X}^{(3)}_{\{2,3,4\}})|\cr
\leq& \frac{1}{2}(|v(\tilde{Z}^{(3)}_{\{2,3,4\}})|+|v(\tilde{X}^{(3)}_{\{2,3,4\}})|)\cr
\leq& \frac{1}{2}(\frac{1}{2}+\frac{1}{2})\cr
=&\frac{1}{2},\cr
    v(\tilde{X}^{(4)})\leq &|v(\tilde{X}^{(4)})|\cr
\leq& \frac{1}{2}|v(Z_1)v(\tilde{X}^{(3)}_{\{2,3,4\}})-v(X_1)v(\tilde{Z}^{(3)}_{\{2,3,4\}})|\cr
\leq& \frac{1}{2}(|v(\tilde{X}^{(3)}_{\{2,3,4\}})|+|v(\tilde{Z}^{(3)}_{\{2,3,4\}})|)\cr
\leq& \frac{1}{2}(\frac{1}{2}+\frac{1}{2})\cr
=&\frac{1}{2}.
\end{align*}

Based on the $3$- and $4$-qubit  scenarios, we conjecture that $v(\tilde{Z}^{(n)})\leq 1/2$ may hold for any $n$-qubit ($n\geq3$) scenario.
Next we shall prove this conjecture by mathematical induction. Since it is true for the $3$- and $4$-qubit  scenarios, then if we assume that it holds true for the $n$-qubit scenario, the key step is to prove that it also holds for the $(n+2)$-qubit scenario.

Note that each term (stabilizer of the associated $n$-qubit state) in the expression of $\tilde{Z}^{(n)}$ is a tensor product of Pauli operators (here only two settings $X$ and $Z$ are involved). Namely, the $2^{n-1}$ terms can be grouped into two classes: $\{S_i=P_{n-1}(i)\otimes Z_n|i=1,2,\cdots 2^{n-2}\}$ and $\{S_j=P_{n-1}(j)\otimes X_n|j=2^{n-2}+1,2^{n-2}+2,\cdots,2^{n-1}\}$, where $P_{n-1}(i)$ and $P_{n-1}(j)$ are  tensor products of $n-1$ Pauli operators (on qubits $1,2,\cdots,n-1$). For any value assignment to the Pauli operators in $P_{n-1}(i)$ (or $P_{n-1}(j)$), denote by $v(P_{n-1}(i))$ (or $v(P_{n-1}(j))$) the product of their values. Clearly, we have
\begin{align*}
   v(P_{n-1}(i))=\pm1,~v(P_{n-1}(j))=\pm1.
\end{align*}
Note that
\begin{align}\label{Zn-ZX-rep}
  \tilde{Z}^{(n)}=&\frac{1}{2^{n-1}}(\sum_{i=1}^{2^{n-2}}P_{n-1}(i)\otimes Z_n\cr
  &+\sum_{j=2^{n-2}+1}^{2^{n-1}}P_{n-1}(j)\otimes X_n).
\end{align}
Therefore,
\begin{align}\label{Zn-plus-2-ZX-rep}
  \tilde{Z}^{(n+2)}=&\frac{1}{2^{n-1}}(\sum_{i=1}^{2^{n-2}}P_{n-1}(i)\tilde{Z}_{\{n,n+1,n+2\}}^{(3)}\cr
  &+\sum_{j=2^{n-2}+1}^{2^{n-1}}P_{n-1}(j)\tilde{X}_{\{n,n+1,n+2\}}^{(3)}).
\end{align}

Similar to the above argument of the $4$-qubit scenario, one have
\begin{align*}
    v(\tilde{Z}^{(n+2)})\leq &|v(\tilde{Z}^{(n+2)})|\cr
\leq& \frac{1}{2^{n-1}}|\sum_{i=1}^{2^{n-2}}v(P_{n-1}(i))v(\tilde{Z}_{\{n,n+1,n+2\}}^{(3)})\cr
  &+\sum_{j=2^{n-2}+1}^{2^{n-1}}v(P_{n-1}(j))v(\tilde{X}_{\{n,n+1,n+2\}}^{(3)})|\cr
  \leq& \frac{1}{2^{n-1}}(\sum_{i=1}^{2^{n-2}}|v(P_{n-1}(i))v(\tilde{Z}_{\{n,n+1,n+2\}}^{(3)})|\cr
  &+\sum_{j=2^{n-2}+1}^{2^{n-1}}|v(P_{n-1}(j))v(\tilde{X}_{\{n,n+1,n+2\}}^{(3)})|)\cr
    \leq& \frac{1}{2^{n-1}}(2^{n-2}|v(\tilde{Z}_{\{n,n+1,n+2\}}^{(3)})|\cr
  &+2^{n-2}|v(\tilde{X}_{\{n,n+1,n+2\}}^{(3)})|)\cr
  \leq& \frac{1}{2}(\frac{1}{2}+\frac{1}{2})\cr
=&\frac{1}{2}.\cr
\end{align*}

Likewise, one can also prove that $v(\tilde{X}^{(n)})\leq 1/2$, where $n\geq3$.

Following the steps given by Eq.(\ref{Bell-sequence}) in the main text, one can construct  the $n$-qubit ($n\geq3$) Mermin inequality and the Svetlichny inequality. One can choose the following Bell operators
\begin{align*}
    \mB^e_{n,\text{Mermin}}(L)&=2^{n-1}\tilde{Z}^{(n)},\cr
    \mB^e_{n,\text{Svetlichny}}(L)=&2^{n-1}\sqrt{2}\frac{\tilde{X}^{(n)}+\tilde{Z}^{(n)}}{\sqrt{2}},
\end{align*}
Writing them in terms of  normal Pauli operators, one can get the corresponding
$\mB^e_{n,\text{Mermin}}(S)$ and $\mB^e_{n,\text{Svetlichny}}(S)$. By invoking $v(\tilde{Z}^{(n)})\leq 1/2$ and $v(\tilde{X}^{(n)})\leq 1/2$, one can get
\begin{align*}
    \langle\mB_{n,\text{Mermin}}\rangle_c\leq&2^{n-2},\cr
    \langle\mB_{n,\text{Svetlichny}}\rangle_c&\leq2^{n-1}.
\end{align*}
They are the $n$-qubit  Mermin inequality and the Svetlichny inequality respectively.
By contrast,  their quantum upper bounds are
\begin{align*}
    \langle\mB_{n,\text{Mermin}}\rangle_{\max}=&2^{n-1},\cr
    \langle\mB_{n,\text{Svetlichny}}\rangle_{\max}&=2^{n-1}\sqrt{2},
\end{align*}
which can be derived by using the sum of square method.

\subsection{Appendix F: ~The proof of the Theorem in the main text}

First, we introduce a simple lemma.

{\it Lemma.} --- Let $A_1=\vec{a}_1\cdot\vec{\sigma},A_2=\vec{a}_2\cdot\vec{\sigma}$ ($\vec{a}_1$ and $\vec{a}_2$ are unit vectors, and
$\vec{a}_1\times\vec{a}_2\neq\vec{0}$), then for any qubit $\rho$, one can get
\begin{align}\label{uncertainty-qubit-2-Pauli-General}
   \frac{\langle A_1+A_2\rangle^2_{\rho}}{|\vec{a}_1+\vec{a}_2|^2}+\frac{\langle A_1-A_2\rangle^2_{\rho}}{|\vec{a}_1-\vec{a}_2|^2}\leq1,
\end{align}
where  for a $2\times2$ matrix $M$, $\langle M\rangle_{\rho}:=\text{Tr}(\rho M)$,  and $\vec{a}_i\cdot\vec{\sigma}=a_i^xX+a_i^yY+a_i^zZ$ ($i\in\{1,2\}$).

{\it Proof.} --- Let $A_+=\frac{A_1+A_2}{|\vec{a}_1+\vec{a}_2|}$, $A_-=\frac{A_1-A_2}{|\vec{a}_1-\vec{a}_2|}$,
and $A_{\bot}=(\frac{\vec{a}_1+\vec{a}_2}{|\vec{a}_1+\vec{a}_2|}\times\frac{\vec{a}_1-\vec{a}_2}{|\vec{a}_1-\vec{a}_2|})\cdot\vec{\sigma}$.
Clearly, we have $A_+^2=A_-^2=A_{\bot}^2=I$ and $\{A_m,A_n\}=i\varepsilon_{ijk}A_k$ ($m,n,k\in\{+,-,\bot\}$). Recall that  a qubit has a standard Bloch vector representation, i.e., $\rho=\frac{I+\vec{n}\cdot\vec{\sigma}}{2}$, where $\vec{n}$ is a unit vector and
$n_x=\langle X\rangle_{\rho},n_y=\langle Y\rangle_{\rho},n_z=\langle Z\rangle_{\rho}$, its two eigenvalues are $\frac{1\pm\sqrt{n_x^2+n_y^2+n_z^2}}{2}=
\frac{1\pm\sqrt{\langle X\rangle_{\rho}^2+\langle Y\rangle_{\rho}^2+\langle Z\rangle_{\rho}^2}}{2}$. Note that $\rho$ is a (semi-) positive operator, its eigenvalues $\frac{1\pm\sqrt{\langle X\rangle_{\rho}^2+\langle Y\rangle_{\rho}^2+\langle Z\rangle_{\rho}^2}}{2}\geq0$. Thus
we have $\langle X\rangle_{\rho}^2+\langle Y\rangle_{\rho}^2+\langle Z\rangle_{\rho}^2\leq1$. Likewise, $\rho$ can also be written as
$\rho=\frac{I+\text{tr}(\rho A_+)A_++\text{tr}(\rho A_-)A_-+\text{tr}(\rho A_{\bot})A_{\bot}}{2}$. Accordingly, one can get
$\langle A_+\rangle_{\rho}^2+\langle A_-\rangle_{\rho}^2+\langle A_{\bot}\rangle_{\rho}^2\leq1$. Therefore, one also have
$\langle A_+\rangle_{\rho}^2+\langle A_-\rangle_{\rho}^2\leq1$. \hfill $\blacksquare$

{\it Observation 2.} --- Let $|\tilde{0}\rangle$ and $|\tilde{1}\rangle$ be two Bell states defined by Eq.(\ref{logical-qubit-basis}) of the main text,
 and $|\tilde{2}\rangle=(|00\rangle-|11\rangle)/\sqrt{2}$,$|\tilde{3}\rangle=(|01\rangle+|10\rangle)/\sqrt{2}$ be the others, then one have
\begin{align}\label{X-Z-other-Bell-state}
\tilde{Z}|\tilde{2}\rangle=\tilde{Z}|\tilde{3}\rangle=\tilde{X}|\tilde{2}\rangle
=\tilde{X}|\tilde{3}\rangle=\tilde{Y}|\tilde{2}\rangle=\tilde{Y}|\tilde{3}\rangle=\vec{0}.
\end{align}
This property arises directly from the definitions of $\tilde{Z},\tilde{X},\tilde{Y}$ in Eq.(\ref{logical-Pauli-XYZ}) of the main text. Similar results can be derived
in other multi-qubit scenarios.

By invoking the lemma and the above observation, one can prove the following theorem.

{\it Theorem.  --- } Let $\tilde{X}$ and $\tilde{Z}$ be two pseudo Pauli operators defined in Eq.(\ref{logical-Pauli-XYZ}) of the main text.
 Consider two experimental Bell operators $\mB_{\vec{n}_1}^{\text{e}}$ and $\mB_{\vec{n}_2}^{\text{e}}$,  where $\vec{n}_i=(\sin\theta_i,\cos\theta_i)$
  is a unit vector in the $xz$ plane ($i=1,2$), and $\mB_{\vec{n}_i}^{\text{e}}=2\sqrt{2}(\sin\theta_i\tilde{X}+\cos\theta_i\tilde{Z})$. Then
$\mB_{\vec{n}_1}^{\text{e}}$ and $\mB_{\vec{n}_2}^{\text{e}}$ satisfy the following uncertainty relation:
\begin{align}\label{uncertainty-Bell-operator}
\frac{(\langle\mB_{\vec{n}_1}^{\text{e}}\rangle+\langle\mB_{\vec{n}_2}^{\text{e}}\rangle)^2}{|\vec{n}_1+\vec{n}_2|^2}
+\frac{(\langle\mB_{\vec{n}_1}^{\text{e}}\rangle
-\langle\mB_{\vec{n}_2}^{\text{e}}\rangle)^2}{|\vec{n}_1-\vec{n}_2|^2}\leq8.
\end{align}

{\it Proof.} ---
Note that  any two-qubit pure state can be expressed as
$|\tilde{\phi}\rangle=\alpha_0|\tilde{0}\rangle+\alpha_1|\tilde{1}\rangle+\alpha_2|\tilde{2}\rangle+\alpha_3|\tilde{3}\rangle$.
Let $|\tilde{\phi}_k\rangle=\alpha_0^k|\tilde{0}\rangle+\alpha_1^k|\tilde{1}\rangle+\alpha_2^k|\tilde{2}\rangle+\alpha_3^k|\tilde{3}\rangle$, and
$|\tilde{\phi}_k^{\prime}\rangle=(\alpha_0^k|\tilde{0}\rangle+\alpha_1^k|\tilde{1}\rangle)/\sqrt{|\alpha_0^k|^2+|\alpha_1^k|^2}$
 if $|\alpha_0^k|^2+|\alpha_1^k|^2\neq0$; otherwise $|\tilde{\phi}_k^{\prime}\rangle=\vec{0}$ (NOT a quantum state).
A general two-qubit system can be represented as $\tilde{\rho}=\sum_{k\in\mI}p_k|\tilde{\phi}_k\rangle\langle\tilde{\phi}_k|$,
 where $\mI$ is an index set, $\sum_{k\in\mI}p_k=1$ and $\forall k,~p_k\geq0$.

If $|\alpha_0^k|^2+|\alpha_1^k|^2=0$ holds for any $k$, then according to Eq.(\ref{X-Z-other-Bell-state}) from the main text,
one can get $|\langle \mB_{\vec{n}}^{\text{e}} \rangle_{\tilde{\rho}}|=0$, i.e., the inequality  to be proved also holds true.
Otherwise, denote $\tilde{\varrho}^{\prime}=\sum_{k\in\mI^{\prime}}p_k(|\alpha_0^k|^2+|\alpha_1^k|^2)|\tilde{\phi}^{\prime}_k\rangle\langle\tilde{\phi}^{\prime}_k|$
and its normalization as
$\tilde{\rho}^{\prime}=\frac{\tilde{\varrho}^{\prime}}
{\sum_{k\in\mI^{\prime}}p_k(|\alpha_0^k|^2+|\alpha_1^k|^2)}$,
where $\mI^{\prime}=\{k|k\in\mI,~|\tilde{\phi}^{\prime}_k\rangle\langle\tilde{\phi}^{\prime}_k|\neq0\}$ and note that
 $\sum_{k\in\mI^{\prime}}p_k(|\alpha_0^k|^2+|\alpha_1^k|^2)\leq\sum_{k\in\mI^{\prime}}p_k\leq1$.
Let $\vec{m}=(\sin\theta,0,\cos\theta)$, i.e., $\vec{m}|_{xz}=\vec{n}$, then $\mB_{\vec{n}}^{\text{e}}=2\sqrt{2}\vec{m}\cdot\vec{\tilde{\sigma}}$.
 By invoking Eq.(\ref{X-Z-other-Bell-state}) of the main text, one can get
$|\langle \vec{m}\cdot\vec{\tilde{\sigma}}\rangle|\equiv|\langle \vec{m}\cdot\vec{\tilde{\sigma}}\rangle_{\tilde{\rho}}|
=\frac{1}{2\sqrt{2}}|\text{Tr}(\mB_{\vec{n}}^{\text{e}}\tilde{\rho})|=\frac{1}{2\sqrt{2}}|\text{Tr}(\mB_{\vec{n}}^{\text{e}}\tilde{\varrho}^{\prime})|\leq
\frac{1}{2\sqrt{2}}|\text{Tr}(\mB_{\vec{n}}^{\text{e}}\tilde{\rho}^{\prime})|$, or other equivalent representations:
\begin{align}\label{General-Bell-reduced-expectation}
  |\langle \vec{m}\cdot\vec{\tilde{\sigma}}\rangle|
  \leq|\langle \vec{m}\cdot\vec{\tilde{\sigma}}\rangle_{\tilde{\rho}^{\prime}}|;~~
  |\langle \mB_{\vec{n}}^{\text{e}}\rangle|\leq|\langle \mB_{\vec{n}}^{\text{e}}\rangle_{\tilde{\rho}^{\prime}}|.
\end{align}

Likewise, let $\vec{m}_i=(\sin\theta_i,0,\cos\theta_i)$, one have $\vec{m}_i|_{xz}=\vec{n}_i$ and $\mB_{\vec{n}_i}^{\text{e}}
=2\sqrt{2}(\sin\theta_i\tilde{X}+\cos\theta_i\tilde{Z})=2\sqrt{2}\vec{m}_i\cdot\vec{\tilde{\sigma}}$. By invoking Eq.(\ref{General-Bell-reduced-expectation}), we can obtain
$\frac{(\langle\mB_{\vec{n}_1}^{\text{e}}\rangle\pm\langle\mB_{\vec{n}_2}^{\text{e}}\rangle)^2}{|\vec{n}_1\pm\vec{n}_2|^2}
=8\langle\frac{\vec{m}_1\pm\vec{m}_2}{|\vec{m}_1\pm\vec{m}_2|}\cdot\vec{\tilde{\sigma}}\rangle^2
\leq8\langle\frac{\vec{m}_1\pm\vec{m}_2}{|\vec{m}_1\pm\vec{m}_2|}\cdot\vec{\tilde{\sigma}}\rangle_{\tilde{\rho}^{\prime}}^2$. According to the above lemma,  one have
$\frac{(\langle\mB_{\vec{n}_1}^{\text{e}}\rangle+\langle\mB_{\vec{n}_2}^{\text{e}}\rangle)^2}{|\vec{n}_1+\vec{n}_2|^2}
+\frac{(\langle\mB_{\vec{n}_1}^{\text{e}}\rangle
-\langle\mB_{\vec{n}_2}^{\text{e}}\rangle)^2}{|\vec{n}_1-\vec{n}_2|^2}\leq
8(\langle\frac{\vec{m}_1+\vec{m}_2}{|\vec{m}_1+\vec{m}_2|}\cdot\vec{\tilde{\sigma}}\rangle_{\tilde{\rho}^{\prime}}^2
+\langle\frac{\vec{m}_1-\vec{m}_2}{|\vec{m}_1-\vec{m}_2|}\cdot\vec{\tilde{\sigma}}\rangle_{\tilde{\rho}^{\prime}}^2)
\leq8$. \hfill $\blacksquare$

\end{appendices}


\begin{thebibliography}{34}
\bibitem [{ano()}]{anotherreplacement}%
  \BibitemOpen
  \href@noop {} {}\bibinfo {note} {Another way to get the Nagata-Koashi-Imoto
  quadratic Bell inequality: One can still use the Bell operators invoked in
  the construction of Uffink's quadratic Bell inequality, but replace
  $Z_1,X_1,Z_2,X_2$ with $A_1, B_1,(A_2+B_2)/\sqrt{2},(A_2-B_2)/\sqrt{2}$
  respectively.}\BibitemShut {Stop}%
\end{thebibliography}
\end{document}